\documentclass[12pt]{article}
\usepackage{epsfig}
\usepackage{amsmath}

\newcommand{\be}{\begin{equation}}
\newcommand{\ee}{\end{equation}}
\newcommand{\beq}{\begin{eqnarray}}
\newcommand{\eeq}{\end{eqnarray}}

\newcommand{\abra}{\left(\!\! \begin{array}{cc} }
\newcommand{\aket}{\end{array}\!\!\right)}

\newcommand{\gb}{\bar\gamma}
\newcommand{\mat}[1]{
  \begin{aligned}
    #1
  \end{aligned}
}

\newcommand{\ben}{\begin{enumerate}}
\newcommand{\een}{\end{enumerate}}
\newcommand{\bit}{\begin{itemize}}
\newcommand{\eit}{\end{itemize}} 
\newcommand{\la}[1]{\label{#1}}

\date{}

\title{\bf Remarks on pole trajectories for resonances}

\vspace{3mm}
\author{C.~Hanhart$^1$, J.R. Pelaez$^2$, G. Rios$^3$ \\ \\
\small $^1$Institut f\"ur Kernphysik, Institute for Advanced Simulations
and J\"ulich Center for Hadron Physics, \\
\small  Forschungszentrum J\"ulich %GmbH,
52425 J\"ulich, Germany; 
\\
\small $^2$ Dept. F\'{\i}sica Te\'orica II. Universidad Complutense, 28040, Madrid, Spain.
\\
\small $^3$ 
Helmholtz-Institut f\"ur Strahlen- und Kernphysik, Universit\"at Bonn, D-53115 Bonn, Germany}

\begin{document}
%\large
\maketitle%\small

\vspace{-10mm}

\begin{abstract}
We discuss in general terms pole trajectories of resonances coupling to a
continuum
channel as some strength
parameter is varied. It is demonstrated that, regardless the underlying
dynamics, the trajectories of poles that couple to the continuum in a
partial
wave higher than $s$--wave are qualitatively the same, while in case of
$s$-waves the pole trajectory can reveal important information on the
internal structure of the resonance. 
In addition we show that only molecular (or extraordinary) states 
appear near thresholds naturally, while more compact structures 
need a significant fine tuning in the parameters. 

This study is of current relevance especially in strong interaction physics, since
lattice QCD may be employed to deduce the pole trajectories for hadronic
resonances as a function of the quark mass thus providing additional, new
access to the structure of $s$--wave resonances. 
\end{abstract}

\section{Introduction}   %% 11111 %%

If all mesons were $\bar qq$ states then there would
be no natural reason for poles in scattering amplitudes to occur very close to
thresholds.  At large values of $N_{c}$, the number of colors in QCD, all
$\bar qq$ mesons become narrow with (nearly) unchanged mass \cite{'tHooft:1973jz}. Thus, their masses have
no relation to the masses of the mesons to which they couple\footnote{The  same is true for a straightforward extension of 
tetraquarks to large $N_c$ \cite{Weinberg:2013cfa} --- in case they existed at large $N_c$
\cite{Cohen:2014tga} --- although other possible extensions of tetraquarks to $N_c\neq3$ lead
to masses that grow when $N_c$ is increased \cite{Cohen:2014vta}.}. 
Accordingly,  the $\rho$ mass is not related to $2m_{\pi}$, nor is the $K^{*}$ mass
related to $m_{K}+m_{\pi}$. On the contrary, there is good reason for ``extraordinary'' hadrons, 
often called hadronic molecules, to have masses close to
thresholds~\cite{Jaffe:2007id}. So the mere fact that the $f_{0}(980)$ and
$a_{0}(980)$ appear very near $K\bar K$ threshold is already a reason to be
suspicious that they may not be simple $\bar qq$ states.  The same applies to
unusual charmonium states that have been found near charm-anticharm meson
thresholds like the famous $X(3872)$ located very close to the $D^0\bar D^{0 *}$ threshold --- for a recent review see~\cite{hqwg}.

  In this paper we look carefully at the way
that the manifestations of poles in scattering amplitudes change as the poles
approach thresholds as some strength parameter is varied --- here one may
think of varying the quark masses in lattice QCD calculations. This has acquired 
a renewed interest after the trajectory  
of the $\sigma$ or $f_0(500)$ resonance pole as a function of the quark mass
was predicted by us
within unitarized Chiral Perturbation Theory \cite{Hanhart:2008mx}.
 A similar trajectory as that of the $\sigma$ was soon shown to be followed by the controversial
$\kappa$ or $K(800)$ resonance in the isospin 1/2 scalar $\pi K$ scattering partial wave, 
including the appearance of a virtual states at sufficiently large pion masses \cite{Nebreda:2010wv}. 
Recently the existence of such a virtual bound state in $\pi K$ scattering at high pion masses has been confirmed by  lattice 
calculations \cite{Dudek:2014qha}. The subtleties of the extraction of resonance parameters from lattice QCD simulations performed at
a finite volume are outlined in detail in Refs.~\cite{Bernard:2010fp,Doring:2011vk} and
will not be discussed further here.

While finishing this work, we became aware of
 a theoretical study \cite{Hyodo:2014bda} of the scaling of hadron masses
near an $s$-wave threshold, showing that the bound state energy is not continuously  connected 
to the real part of the resonance energy. In this paper we have another look at this issue
which allows us to provide various additional, non-trivial insights.
 In particular, we
demonstrate that there is a qualitative difference between the pole
trajectories of resonances that couple to the relevant continuum channel is an
$s$--wave or in a higher partial wave: As a consequence of analyticity a
resonance is characterized by two poles on the second sheet, one located at
$s=s_R$ and one located at $s=s_R^*$. For narrow resonances only one of them
is close to the physical region. As some strength parameter is increased, the
two poles start to approach each other. We will show on general grounds
that while for higher partial waves the poles meet at the corresponding
two meson threshold, for $s$--waves the poles can still be located inside the
complex plane even for the real part of the pole position at or below
threshold. 
As a consequence, $s$--wave--trajectories are controlled by an
additional dimensionful parameter, namely the value of $s$ where the two poles
meet below threshold which may be related to the structure of the state.
In other words, generic trajectories of $s$-wave resonances
do lead to poles 
whose real part of the position is below threshold, but whose imaginary part of the position does not vanish,
before giving rise to virtual bound states, and then bound states, as some strength parameter is varied. 
While this observation is in line with the findings of Refs.~\cite{Hanhart:2008mx,Guo:2009ct},  it
is in vast conflict with
``common wisdom'' that the imaginary part of a pole has to be identified with one half of its decaying width,
for this implies that, if the ``resonance mass'' --- identified with the real part of the pole position ---
 lies below threshold, the pole necessarily has to 
lie on the real axis. Of course, such identification is a fair approximation 
for narrow resonances far from thresholds or other singularities, but very inappropriate
in the cases we will show below.

The paper is organized as follows: in the next Section we discuss general properties of the 
poles that appear in the $S$-matrix, paying especial attention to poles that occur
in partial waves with angular momenta higher than 0, and in particular to the role of the centrifugal barrier which is absent in
the scalar partial waves. Next we consider the
trajectories of resonance poles in the complex plane as a function of some strength parameter,
and how they can become bound states. In the next section we briefly review Weinberg's compositeness criterion
and reformulate it in terms of the parameters introduced in the previous section.
The possible behaviors are then illustrated with two models of scattering in separable potentials 
within non-relativistic scattering theory, one with a single channel and another one in a two-channel system.
In Section 4 we analyze 
the realistic examples of the pole trajectories of the $\sigma$
or $f_0(500)$ scalar meson and the $\rho(770)$ as functions of the quark masses, obtained
from the combination of Chiral Perturbation Theory and a single channel dispersion relation
obtained in \cite{Hanhart:2008mx}.
We show how 
the generic features discussed in this paper show up in these two cases. 
In particular, we can conclude that the $f_0(500)$ or sigma meson would have a predominantly
molecular nature, if the pion mass were of the order of $450$ MeV or higher.
In the final Section we summarize our results.

\section{General properties of $S$--matrix poles}

In this work we only consider one continuum channel.  This implies
that the $S$--matrix has one right hand cut, starting at
$s=(2m)^2$ --- the so
called unitarity cut\footnote{For simplicity we only consider the case of
  scattering of two particles with equal mass, however, the
  generalization to unequal masses is straightforward.}. As a consequence there are two sheets
  and, as usual, we call first or physical sheet the one
corresponding to a momentum with a positive imaginary part.  The $S$
matrix evaluated on sheet $I$ ($II$) is written as $S_I(s)$
($S_{II}(s)$). If no subscript is given, the expression holds for both
sheets.  It follows directly from unitarity and analyticity
that~\cite{book}
\begin{equation}
S_I(s)=1/S_{II}(s) \quad \mbox{and} \quad \left[S(s)\right]^*=S(s^*) \ .
\label{sprops}
\end{equation}
As a consequence a pole on the second sheet immediately implies a zero
on the first and vice-versa. In addition, if there is a pole at $s=s_0$, there must
also be a pole at $s=s_0^*$, i.e., poles outside the real axis occur in conjugate pairs.
Furthermore, it can be shown that the only poles 
allowed on the physical sheet are bound state poles, namely, those located
on the real axis below threshold.

\begin{figure}[t!]
\begin{center}
\begin{center}
\epsfig{file=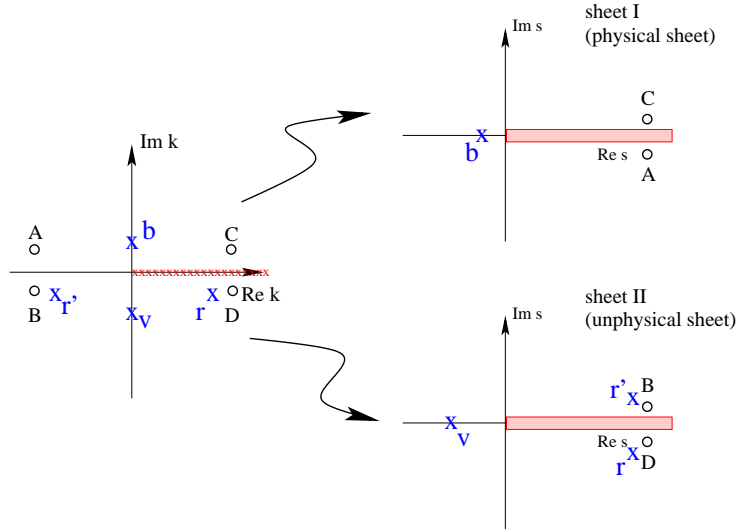,width=0.6\textwidth}
\end{center}
\caption{Relation between $k$--plane and $s$--plane:
on the left the $k$--plane is shown. The (red) $x$s denote the 
physical axis. On the right the two $s$--plane sheets are shown.
Here the broad band indicates the position of the unitarity cut.
The upper (lower) half plane of the $k$--plane
maps onto the first (second) sheet in the $s$--plane such that 
 the points $A$-$D$ get transferred as indicated in the figure.
In addition, the allowed pole positions in the complex plane
are also shown as $x$. They are labeled as $b$ for the 
bound state, $v$ for the virtual state, and $r$ and $r'$ for
the two conjugate poles of the resonance state.}
\label{kvss}
\end{center}
\end{figure}

A different, but equivalent, way to discuss the pole structure of the
$S$--matrix is to use the $k$--plane: instead of the Mandelstam
variable $s$, the center of mass momentum $k$ is used to characterize
the energy of the system. The two quantities are related via
\begin{equation}
k = \sqrt{s/4-m^2} \ .
\end{equation}
The obvious advantage is that there is no right hand cut with respect
to $k$ and correspondingly there is only one sheet.  It follows
directly from the definition that the upper (lower) half plane of the
complex $k$--plane, defined by positive (negative) values of the
imaginary part of $k$, maps onto the first (second) sheet of the
$s$--plane.  The conditions derived above from Eq.~(\ref{sprops}) translate
into the $k$--plane as follows: the only poles allowed in the upper half plane are
on the imaginary axis and in the lower half plane appear as mirror
images with respect to the imaginary axis. 
The relation between the different planes is illustrated in Fig.~\ref{kvss}.
On the one hand, it becomes clear from the figure that the resonance pole located at $r$
 is the one closest to the physical axis and therefore physically more
relevant than the one at $r'$ in the vicinity of the pole. On the other hand, 
near threshold both poles are equally relevant regardless where they are located
in the second sheet.
Finally, in the $k$ plane virtual states appear as poles on the negative
imaginary axis (labeled as $v$ in the figure) and bound states
as poles on the positive imaginary axis (labeled as $b$ in the figure).

Now, assuming that there is at least one resonance pole, and that it is not too far away from threshold,
we are now in the position of writing down  the most general expression
for the $S$--matrix in the vicinity of that pole or its conjugate partner.
For the derivation it is easier to use
 the $k$ plane, and thus we assume that there is a resonance pole at
$k=k_{\rm p}-i\gamma$ with $\gamma>0$.
For a resonance, $k_{\rm p}$ is a real number and we choose $k_{\rm p}>0$, for, as commented above,
it corresponds to the pole closest to the physical axis. 
Then, from the above considerations it follows that there is in addition
a pole at $k=-k_{\rm p}-i\gamma$ and zeros at $k=\pm k_{\rm p}+i\gamma$. We may
therefore, dropping terms of higher order in $k$, 
and for a particular partial wave ${\ell}$,
write the following general expression for the 
$S$--matrix element in the vicinity of the pole~\cite{book}:
\begin{equation}
S_{\ell}(k) =e^{i\phi(k)}\frac{(k-k_{\rm p}-i\gamma)(k+k_{\rm p}-i\gamma)}{(k-k_{\rm p}+i\gamma)(k+k_{\rm p}+i\gamma)} \ ,
\label{Spara}
\end{equation}
where $\phi(k)$ is a smooth function, real valued for real, positive values of $k$. For simplicity
this phase factor will be dropped in what follows.
Using the definition of the $T$ matrix, $S=1+2ikT$, we may write
\begin{equation}
T_{\ell}(k) =-\frac{2\,\gamma}{k^2-(\gamma^2+k_{\rm p}^2)+2i\gamma k} \ .
\label{Tpara}
\end{equation}

For elastic scattering unitarity provides a stringent link between the real and the
imaginary part of $T$, that actually allows the
$T$--matrix elements to be described in terms of its phase $\delta$ as
\begin{equation}
\arctan (\delta) = -\frac{2\,k\,\gamma}{k^2-(\gamma^2+k_{\rm p}^2)} \ .
\label{phase}
\end{equation}

Of course, it is straightforward to recast the above expressions in terms of $s$ instead of $k$. 
One finds for example
\begin{equation}
S_{\ell}(k) =\frac{s-s_0-4i(s-4m^2)^{1/2}\gamma}{s-s_0+4i(s-4m^2)^{1/2}\gamma} \ ,
\label{Spara_s}
\end{equation}
with $s_0=4\left(k_{\rm p}^2+\gamma^2+m^2\right)$.

\subsection{$\ell>0$ partial wave threshold behavior and poles}

 In general the centrifugal barrier demands, for  momenta
much
 smaller than some typical scale $\mu$,
that the scattering amplitude behaves as $T_{\ell}\propto k^{2\ell}$.
If we are only interested in this low energy region, 
the constraint translates into the replacement
\begin{equation}
\gamma = \gamma(k)=\bar \gamma k^{2\ell} \ .
\label{gol}
\end{equation}
Of course, this amplitude should only be used for $k$ much smaller than
the typical scale $\mu$, not beyond. 
Note that $\ell=0$ waves are unaffected by this change, but, for example, $\ell=1$ waves
now have poles whenever $i\bar \gamma k^2+k\pm k_{\rm p}=0$, namely at:
\begin{equation}
k_{\rm pole}=\frac{i}{2\bar\gamma}
\Big[ 1\pm\sqrt{1\mp4i\bar\gamma k_{\rm p}}\Big],\qquad (\ell=1\;\hbox{case}).
\end{equation}

These are four poles in conjugated pairs, but of course, 
they are only meaningful if
they lie within the low momentum region of validity of our amplitude.
We can ensure that we have only one conjugated pair within this region
if we require $\bar\gamma k_{\rm p}\ll 1$, (which is nothing but assuming that 
both parameters are natural, i.e., $\bar \gamma\ll 1/\mu$ and $k_{\rm p}\ll \mu$).
In such case we can expand
\begin{equation}
  \label{squareroot}
  \sqrt{1\mp 4i\gb k_{\rm p}}\simeq 
%  1 \mp \frac1{2}(4i\gb k_{\rm p}) - \frac1{8}(4i\gb k_{\rm p})^2 = 
  1 \mp 2i\gb k_{\rm p} + 2\gb^2k_{\rm p}^2 + ...,
\end{equation}
so that the four poles lie at:
\begin{equation}
  \label{kp}
  k_{\rm pole}\simeq \frac{i}{2\gb}\Big[1 \pm 
    \left(1 \mp 2i\gb k_{\rm p} + 2\gb^2k_{\rm p}^2 \right)
  \Big]=\left\{\mat{
&
%-\frac{i}{2\gb}\Big[\mp 2i\gb k_{\rm p} + 2\gb^2k_{\rm p}^2\Big]=
\mp k_{\rm p} - i\gb k_{\rm p}^2. \qquad\qquad\hbox{ (Physical pair)}\\ 
& 
%-\frac{i}{2\gb}\Big[-2\pm 2i\gb k_{\rm p} - 2\gb^2k_{\rm p}^2\Big]=
%\pm k_{\rm p} + i\gb\left(\frac1
%{\gb^2}+k_{\rm p}^2\right)
\pm k_{\rm p}+\frac{i}{\bar\gamma}\Big(1+(\gb k_{\rm p})^2\Big)
\hbox{ (Unphysical pair)}
}\right.
\end{equation}
One should not worry about the  unphysical conjugated pair of poles
lying on the first sheet, since our amplitude 
has been constructed for $1/ \bar \gamma\gg k_{\rm p}$
 and thus
these spurious poles are deep in the complex plane, 
beyond the range of applicability of our approach, which is however valid for 
the two poles not too far from threshold. 
A similar pattern emerges for even higher partial waves, with a physical pair
for small $k$ and additional unphysical pairs of poles beyond the applicability 
region of our amplitude.

\subsection{Pole trajectories as a function of a strength parameter}

In the construction presented in the previous section it was assumed that
$k_{\rm p}$ is a real number --- then the equations describe a resonance.  We will
now generalize this investigation by considering the movement of the poles as
some strength parameter is varied. Therefore we will study how the resonance
properties change when varying $k_{\rm p}$.  In particular, it is interesting to
observe the trajectories of the poles for $k_{\rm p}\rightarrow0$ especially very
 close to threshold.  Of course, as long as two
conjugate poles exist, their trajectories have to be symmetric with respect to
the imaginary $k$ axis.

\begin{figure}
\begin{center}
\includegraphics[width=7cm]{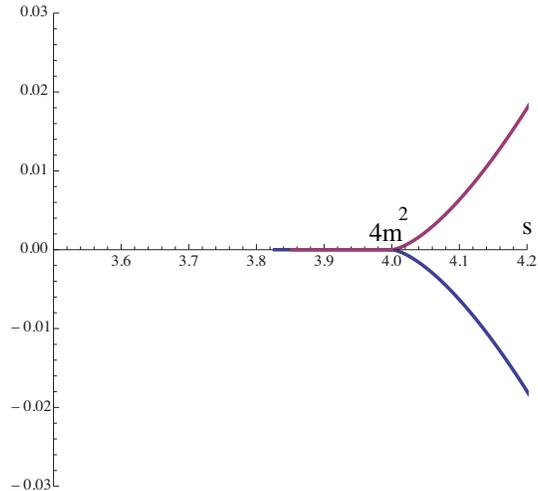}
\vspace*{-0.1in}
\end{center}
\caption[x]{\footnotesize Motion of the p-wave poles in the complex $s$-plane
  with $m=1$ and $\bar \gamma=0.2$.} 
\label{Fig:ppoles}
\end{figure}

Let us first follow the trajectories followed by conjugate poles
for $\ell>0$ partial waves. As a consequence of Eq.~(\ref{gol}), they  will
come infinitesimally close to $k_{\rm pole}=0$. 
However, as commented above 
there can be no poles of the $S$--matrix on the physical cut.
This property is automatically implemented in Eq.~(\ref{Spara})
for when $k_{\rm p}=0$ and $\gamma=0$ simultaneously, the zeros
in the numerator and denominator
cancel to yield $S(k=0)=1$. The resulting pole trajectories in the $s$ plane are illustrated
for $\ell=1$ in Fig.~\ref{Fig:ppoles}.

In contrast,  for $s$--waves 
 the point where the two conjugate poles meet each other on the imaginary $k$ axis
is not fixed except for the condition that no poles in the physical axis 
should exist in the first Riemann sheet, and in particular not at $k=0$. 
But that leaves the whole negative axis
for s-wave poles to meet when $k_{\rm p}$ decreases
and the point where the two trajectories meet,  $-i\gamma$, 
is a non--trivial parameter of the underlying dynamics.
This is one of the central messages of this paper.

In order to extend our discussion to poles
below threshold, we need to continue analytically 
$k_{\rm p}$ to complex values $k_{\rm p}=i\kappa_{\rm p}$, with  $\kappa_{\rm p}$ real and positive
so that $k_{\rm p}^2$ is real and negative. Of course, for our
amplitude to make sense we still keep the condition
$\gb\kappa_{\rm p}\ll1$.

Now, for the $\ell >0$ case, we find two physical  poles at $k_{pole}=i\kappa_{\rm p}(\pm1+\gb \kappa_{\rm p})$.
In the s-plane, these are two poles 
below threshold but one in the first and another one in
the second Riemann sheet. 
The resonance has become a bound state.
Since $\gb\kappa_{\rm p}\ll1$ they lie almost symmetrically
with respect to the threshold. 
As seen from the $s$-plane, this is the 
typical structure of subthreshold poles
in the first and second Riemann sheets. 

\begin{figure}
\begin{center}
\includegraphics[width=7cm]{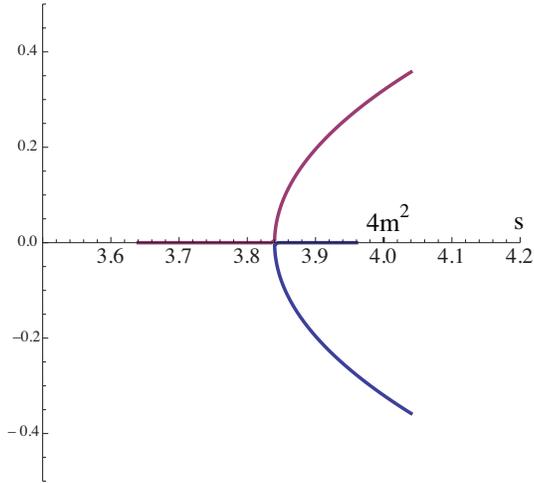}
\vspace*{-0.1in}
\end{center}
\caption[x]{\footnotesize Motion of the s-wave poles in the complex $s$-plane with $m=1$ and $\gamma=0.2$. }
\label{Fig:spoles}
\end{figure}

In contrast, for scalar waves, we 
find two poles in the imaginary axis at $k_{pole}=i(\pm\kappa_{\rm p}-\gamma)$.
Note that, as $\kappa_{\rm p}$ grows, 
the two poles start separating from each other and move apart from
 the ``meeting point'', $-i\gamma$. In the $k$--plane
both move along the imaginary $k$--axis, the physically more relevant one
is located at $-i(\gamma-\kappa_{\rm p})$ while the other one is located at $-i(\gamma+\kappa_{\rm p})$.
Correspondingly in the $s$--plane the two poles move along the 
real axis below threshold, but both of them lying on the second sheet
until $\kappa_{\rm p}=\gamma$. Eventually, when $\kappa_{\rm p}>\gamma$, the first pole moves
to the physical sheet --- the virtual state turns into a bound state.
The corresponding motion of the poles is illustrated in Fig.~\ref{Fig:spoles}.

One may define the mass $M$ of a particle as the real part of 
the corresponding pole position in the complex plane. It is
therefore interesting to follow this point as $\gamma$ and $k_{\rm p}$
vary. For $s$--waves, in general one finds a striking non--analytic behavior
in $M$ at the point where $k_{\rm p}=0$. On the other hand, for partial
waves higher than $s$--waves the behavior much smoother
\footnote{This non-analytical behavior in hadron masses may also propagate to other
  observables, like form factors in the $t$--channel~\cite{Guo:2011gc}.}. 
This non-analyticity of the hadron mass when the conjugate poles reach the real axis
as $k_{\rm p}\to 0$
has been recently studied in detail within the general formalism of Jost functions in \cite{Hyodo:2014bda}. 
The conclusion is a similar warning to the one we raised in Ref.~\cite{Hanhart:2008mx} about  
the naive mass extrapolation formulas for states which appear near thresholds
on the lattice, although within a more general framework. 
In this work we will  illustrate this non-analyticity 
in passing, when explaining the different possible pole trajectories
on the basis of various examples below, whereas for the analytic formalism of those 
mass singularities we simply refer the reader to \cite{Hyodo:2014bda}.

\section{Summary of Weinberg's Criterion}
\label{nature}

In Refs.~\cite{weinberg} Weinberg developed a criterion for compositeness for
bound states that occur in the $s$-wave of scattering amplitudes (under which
circumstances  this formalism
can be generalized to resonances is described in Ref.~\cite{evidence}). 
The starting point is the scattering amplitude near threshold that may be expressed
in terms of the scattering length, $a$, and the effective range, $r$ as\footnote{Note
that sometimes a different sign convention is used for the scattering length.}
\begin{equation}
T_{0}(k)= \frac{1}{k\cot\delta_{0}(k)-ik} = \frac{1}{-1/a+rk^{2}/2 -ik} \ .
\label{e.9}
\end{equation}
Weinberg derived relations between the scattering length, $a$,
the effective range, $r$, and the wave function renormalization constant for
the particle described by the $S$-matrix pole, $Z$,
\begin{align}
a=2\left(\frac{1-Z}{2-Z}\right)R + {\cal O}(1/\beta) \ , \quad 
r= -\left(\frac{Z}{1-Z}\right)R + {\cal O}(1/\beta) \ .
\label{e.7}
\end{align}
Here $\beta$ is the typical momentum scale of the binding interactions ---
 in our case either $\beta\sim  m_{\pi}$ or larger, depending on whether single pion exchange is important in the process ---
 and $R$ is the inverse of the imaginary momentum corresponding to the $S$-matrix pole
\begin{equation}
R = \frac{1}{\kappa} = \sqrt{\frac{1}{2\mu  B  }}
\label{e.8}
\end{equation}
 when the pole in $S$ occurs at $s=4(m^{2}-\kappa^{2})=4m^{2}-4mB$.
  For a
bound state $\kappa>0$ and for a virtual state $\kappa<0$ (in both cases
$B>0$).
On general grounds one can show that  in leading order in an expansion in $1/(R\beta)$ $Z$
can be interpreted as the probability to find the ordinary, compact component in the
wave function of the physical state; especially $0\le Z\le 1$.

Assuming that the effective range approximation is valid for all
momenta of interest, from Eq.~(\ref{e.9}) evaluated at the pole follows a single
kinematic relation among $a$, $r$, and $R$
\begin{equation}
\frac{1}{R}=\frac{1}{a} +\frac{r}{2R^{2}}
\label{e.10}
\end{equation}
Weinberg found that a predominantly composite state (or hadronic molecule
or extraordinary hadron) has  $Z\approx 0$.  In the case of a
weakly bound particle, where $R\gg 1/\beta$, this criterion reduces to $a\sim
R$ and $r\sim 1/\beta$, with the range term in Eq.~(\ref{e.10}) just
providing
a small correction.
Note that within potential scattering one can show that the terms of order
$1/\beta$ are typically positive.

 On the other hand, a predominantly elementary state
has $Z\approx 1$ and therefore $a\sim 0$, or more accurately $a\sim
1/\beta$, and
and $|r|\gg R$. As a result, in order to get a bound state near threshold
for a predominantly elementary state a fine tuning between the range term and the scattering length term is
necessary in Eq.~(\ref{e.10})\footnote{This situation may be accompanied by quite unusual line
shapes as demonstrated in Refs.~\cite{interplay1,interplay2}.}.
This clearly demonstrates, that it is way more natural to find composite
states near thresholds than elementary states.

The argument can be easily expressed in terms of the parameters introduced in
the previous section. From Eq.~(\ref{Tpara}) one finds 
\be
\label{arofgammak}
a=-\frac{2\gamma}{\gamma^2+k_{\rm p}^2} \quad \mbox{and} \quad r=\frac1{\gamma} \ .  
\ee
Using Eq.~(\ref{e.7}) this can be translated to 
\be
\label{Zofgammak}
Z=1-\frac{\gamma}{\kappa_{\rm p}} \ .  
\ee 
As explained in the previous section, for
a shallow bound state one has $\kappa_{\rm p}-\gamma=\kappa\ll \beta$. Thus,
we again recover that the most natural situation for a near threshold
pole is $Z\simeq 0$, since only if both $\kappa_{\rm p}$ and $\gamma$
are individually much smaller than $\beta$ and at the same time $\gamma\ll
\kappa_{\rm p}$, then $Z\simeq 1$, referring
to an elementary state. Clearly, for this to be realized
fine tuning is necessary.

In the following we will illustrate the patterns described above 
on two simple models. (Some pole trajectories within a 
specific coupled channel model were already shown in \cite{Lesniak}).
Both of these models are based on non-relativistic scattering in a separable
potential.  Model A is a single channel separable potential.  If the potential
is attractive and strong enough, it can generate an $S$-matrix pole.  In
\cite{Jaffe:2007id} it was  argued that this is a ``toy model'' for an extraordinary
hadron which would vanish as $N_{c}\to\infty$.  Model B is a two channel model
where there is no diagonal interaction in the open channel, but there is a
bound state in the closed channel (a Feshbach resonance).  In
\cite{Jaffe:2007id} it was argued that this is a model for an ``ordinary hadron'',
whose width would go to zero as $N_{c}\to\infty$.

\subsection{Model A }
Model A has a separable potential that only couples to a single partial wave
with angular momentum $l$.  The scattering amplitude in all other partial
waves is zero.  For the partial wave with angular momentum $l$, the
Schr\"odinger equation is
\be
\la{schr}
-u''_{\ell} (r)+\frac{\ell(\ell+1)}{r^{2}}u_{\ell}(r)- \lambda \int_{0}^{\infty}dr'v(r)v(r')u_{\ell}(r')=Eu_{\ell}(r'),
\ee
To make things simple
 $v(r)$ is chosen such that  the integrals can be done analytically:
\begin{equation}
  \label{eq:V*}
  v(r)=\sqrt{2} \mu^{3/2} e^{-\mu r},
\end{equation}
Although for $r\sim 1/\mu$ the behavior of the system depends
on the form chosen for $v(r)$, the behavior for $r\ll 1/\mu$ is genuine.

Then one finds for the scattering amplitude, $f_{l}(k)$,
\be
f_{l}(k)=\frac{k\lambda \xi_{l}^{2}(k)}{1-\frac{2\lambda}{\pi}\int_{0}^{\infty}dq q^{2}\frac{\xi_{l}^{2}(q)}{q^{2}-k^{2}-i\epsilon}} \equiv\frac{N_l(k)}{D_l(k)}
\la{sepscatt}
\ee
where $\xi_{l}(k)=\int_{0}^{\infty}rj_{l}(kr)v(r)$, and in particular
\begin{equation}
\xi_0(k)=\frac{\sqrt{2}}{k^2+1}.
\label{xinew}
\end{equation}
Here units are chosen such that $\mu=1$ --- accordingly
the model should reproduce the genuine behavior discussed above for $k\ll 1$.
One  can compute $N(k)$ and $D(k)$ explicitly;  for the $s$-wave ($\ell=0$),
\begin{align*}
N_{0}(k)= \frac{2k\lambda}{(k^2+1)^2},\quad
D_{0}(k)= 1+\frac{\lambda}{(k+i)^2}.
\end{align*}

For $s$-wave poles located near a threshold one may use Weinberg's criterion
to pin down the degree of compositeness of the corresponding physical state.
Here closeness to the threshold translates into $k\ll 1$. Then we may
read off from the expressions given above 
\begin{equation}
\label{armodelAnew}
a_A=\frac{2\lambda}{\lambda-1} \quad \mbox{and} \quad
r_A=\frac{\lambda+2}{\lambda}. 
\end{equation}
A  bound state is present  only if the interactions are attractive
and $a>0$, which translates into $\lambda>1$ ($\lambda<0$  refers to a repulsive interaction). In
addition, $r$ is always positive --- which means that in Eq.~(\ref{e.7}) for
the range the $1/\beta$ term dominates.  Thus, the pole is located very near
threshold only for $r_A/(2a_A) \ll 1$, 
 as follows straightforwardly from
Eq.~(\ref{e.10}) --- within the model this ratio does not exceed 0.3 showing
that Model A produces extraordinary hadrons in the whole parameter range
where bound states are produced.
Equivalently one may also directly calculate the wave function renormalization
constant for the $s$-wave --- one finds it consistent with 0 within the
uncertainties. Thus, the single partial wave separable potential
generates an $S$-matrix pole dynamically that mimics an extraordinary hadron: a hadronic molecule.

\subsection{Model B }

This model is designed to show the scattering effects of a confined state when
it can be treated non-relativistically using the Schroedinger equation.
It is described in detail in \cite{Jaffe:2007id},
Section II.C, and will not be  repeated here.  Basically it maps onto a
separable potential model but with $\lambda\to \lambda/(E-E_{0})$, where
$E_{0}$ is the energy of the confined channel bound state.  (Feshbach showed
that near such a state the two channel Schr\"odinger Equation collapses to a single channel
equation with a separable potential.)  Here are the numerator and denominator
of the $s$-wave scattering amplitude for this model:
\begin{align}
N_{0}(k)&= \frac{2k\lambda}{(k^2+1)^2},\\
D_{0}(k)&= k^2-k_0^2+\frac{\lambda}{(k+i)^2}.
\label{feshbachswavenew}
\end{align}

The only difference between the two models is that the $1$ in $D_{0}(k)$ in
model A is replaced by $k^{2}-k_{0}^{2}$ in model B --- said differently:
model A is recovered from model B in the limit $k_0^2\to \infty$ while
$\bar\lambda=-\lambda/k_0^2$ is kept finite. In particular, this implies that in some areas of
parameter space model B describes, as model A, composite states.

However, because of the new extra parameter, $k_{0}$, the location of a near
threshold pole in the scattering amplitude is no longer directly linked to the
scattering length and as a consequence Weinberg's criterion for compositeness
can be evaded.  This is intuitively clear: If the coupling to the confined
state is very weak or repulsive, it should appear like an elementary particle
in the scattering channel. This is the case for very small $\lambda$ and
negative $k_0^2$. If one takes, for instance, $\bar\lambda=0.1$ and 
$k_0=0.3i$ one obtains a bound state with $Z\simeq 1$, which is
an elementary state.

In summary, for certain parameters, that need to be fine--tuned considerably, the ``Feshbach'' resonance model can describe a
bound state in a confined channel that couples to scattering in an open
channel that does not satisfy Weinberg' criterion for compositeness and thus
should be interpreted as a genuine (``ordinary'') state.

\section{A realistic example: Pole trajectories of the $\sigma$ and $\rho$ mesons
as a function of quark masses
}

\begin{figure}%[t!]
\begin{center}
\epsfig{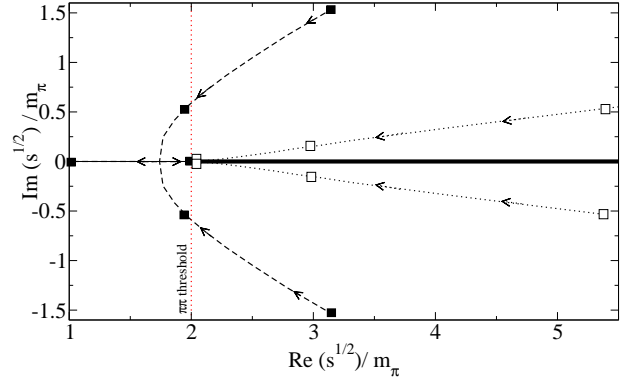}
\caption{ Movement of the $\sigma$ (dashed lines) and $\rho$ (dotted
  lines) poles for increasing pion masses (direction indicated by the
  arrows) on the second sheet as extracted from the IAM.  The filled (open) boxes denote the
  pole positions for the $\sigma$ ($\rho$) at pion masses $m_\pi=1,\
  2,$ and $3 \times m_\pi^{\rm phys}$, respectively. Note, for
  $m_\pi=3m_\pi^{\rm phys}$ three poles accumulate in the plot very near the
  $\pi\pi$ threshold.}
\label{polosNormUpDown}
\end{center}
\end{figure}

\begin{figure}[t!]
  \centering
  \hbox{
    \centering
    \hspace{-1.9cm}
    \includegraphics[scale=.9]{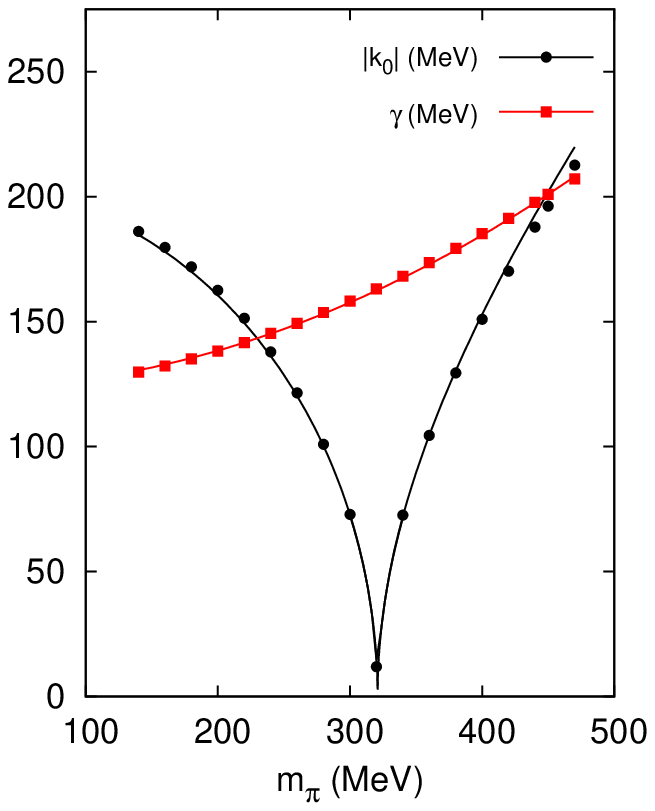}
    \hspace{-4.5cm}
    \includegraphics[scale=.9]{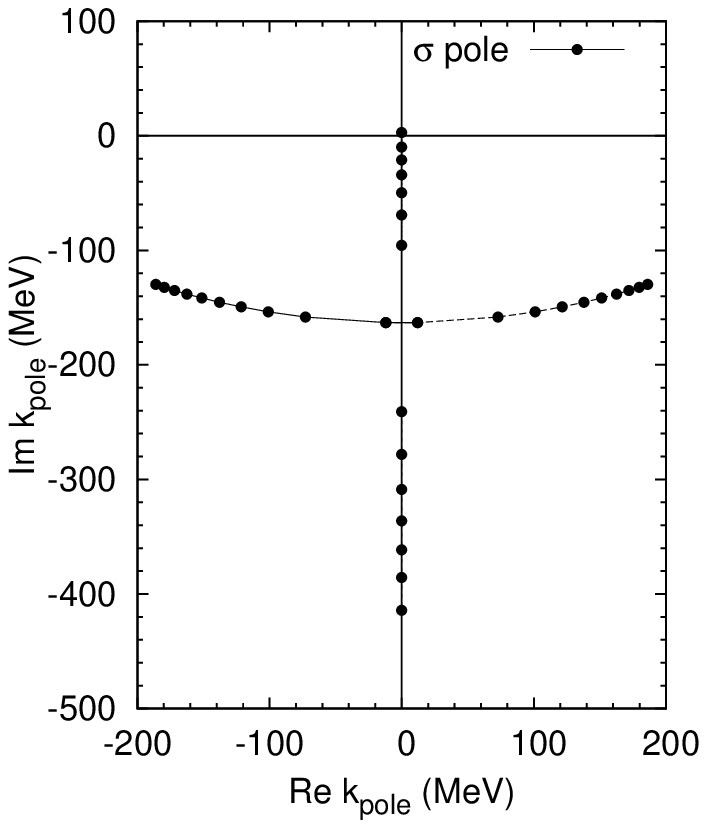}
  }
  \caption{Behavior of the $\sigma$ pole in the
    $k$--plane. Left panel: $m_\pi$ dependence of $k_{\rm p}$ and $\gamma$.
    The filled circles (boxes) show the results of
    the numerical determination for $|k_p|$ ($\gamma$) from
    the full calculation, while the lines are produced from
    the fitting functions given in the text.
    Right panel: the resulting pole movement for the $\sigma$
    in the $k$--plane}
  \label{sigink}
\end{figure}
\begin{figure}[t!]
  \centering
  \hbox{
    \centering
    \hspace{-1.9cm}
    \includegraphics[scale=.9]{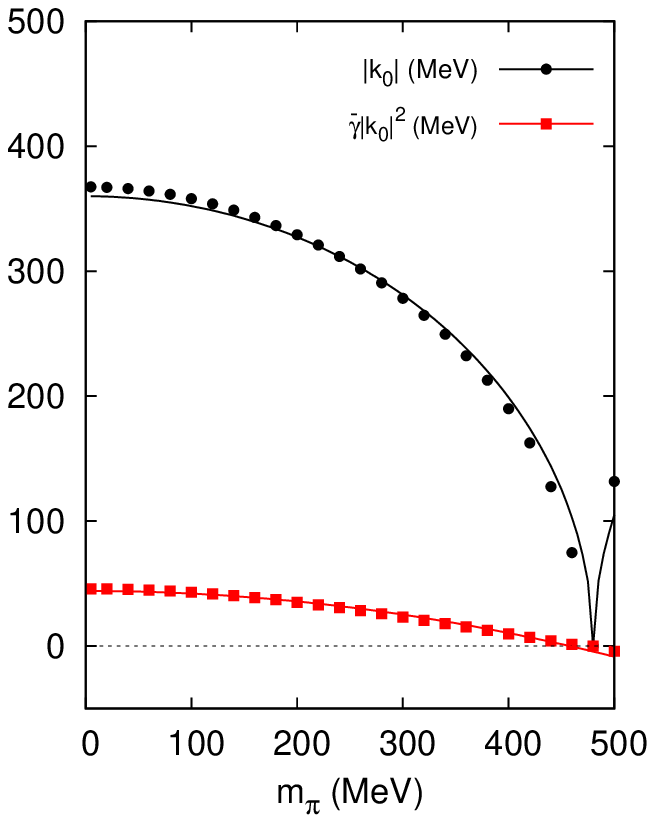}
    \hspace{-4.5cm}
    \includegraphics[scale=.9]{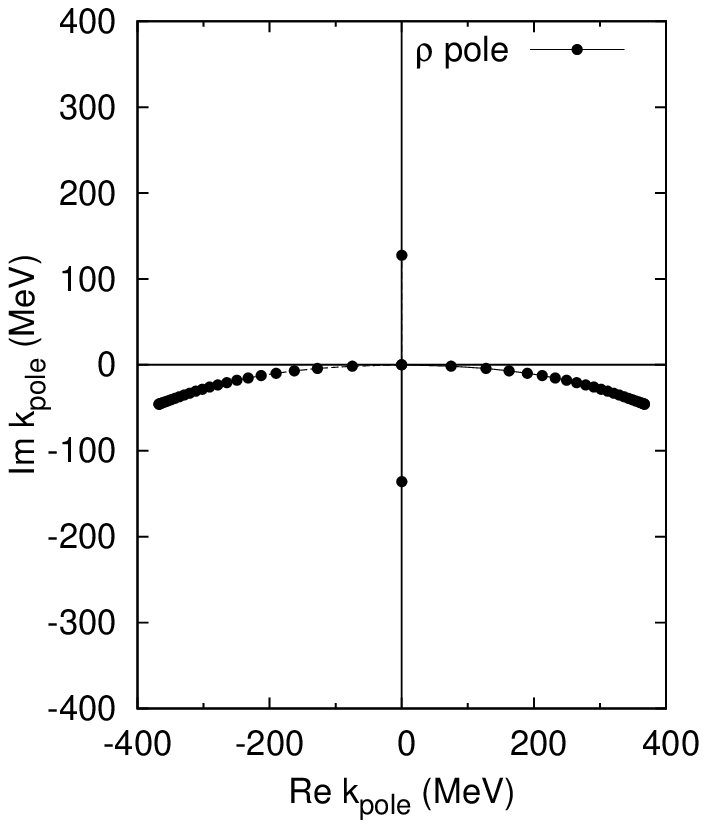}
  }
  \caption{Behavior of the $\rho$ pole in the
    $k$--plane. Left panel: $m_\pi$ dependence of $k_{\rm p}$ and $\bar\gamma$.
    The filled circles (boxes) show the results of
    the numerical determination for $|k_p|$ ($\bar gamma k_p^2$) from
    the full calculation, while the lines are produced from
    the fitting functions given in the text.
    Right panel: the resulting pole movement for the $\rho$
    in the $k$--plane}
  \label{rhoink}
\end{figure}

In order to illustrate with  a realistic example
what was described in the previous
sections, we now show the results
for the pole trajectories of the $\rho$--meson and the $\sigma$--meson
calculated within the inverse amplitude method (IAM)~\cite{IAM}. The approach
uses Chiral Perturbation Theory (ChPT) predictions to a given order to fix the subtraction constants
 of an elastic partial wave dispersion relation. This leads to
an amplitude consistent with elastic unitarity that by construction matches the ChPT
expansion when re-expanded at low energies and at the same time generates the poles associated to the $\sigma$ and $\rho$
resonances in pion-pion scattering~\cite{Dobado:1996ps}. Note that the numerical values of the
low energy constants obtained when fitting the IAM to scattering data
might slightly differ from those of ChPT since
they absorb higher order effects.
Since the whole QCD quark mass dependence is included up to the desired
order in terms of the ChPT expansion of the pion mass and decay constant, 
one can study the quark mass dependence of both the $\sigma$ and $\rho$
resonances \cite{Hanhart:2008mx}. We will now discuss the resulting pole
trajectories is some more detail.

In Fig.~\ref{polosNormUpDown} we show the pole movement in the second sheet
for both $\sigma$ and $\rho$.
The pole movement of the $\sigma$ in the $k$--plane is shown in 
the right panel of Fig.~\ref{sigink}.
Not only provides us the $k$--plane with a different look at the
positions and movements of $S$--matrix poles, it also allows us to
give a simple parameterization for the $m_\pi$--dependence of the
resonance poles shown in Figs.~\ref{polosNormUpDown} and ~\ref{sigink}.
Especially we get for the $\sigma$
\begin{equation}
\left(k_p^\sigma\right)^2=a_\sigma^2(b_\sigma^2-m_\pi^2) \quad \mbox{and} \quad \gamma^\sigma = \gamma_0^\sigma + c_\sigma(m_\pi/m_\pi^{\rm phys.})^2 \ ,
\end{equation}

\noindent
with $a_\sigma= 0.64$ MeV, $b_\sigma=320.8$ MeV, $c_\sigma=7.5$ MeV and $\gamma_0^\sigma=123$ MeV and analogously for the $\rho$
\begin{equation}
\left(k_p^\rho\right)^2=a_\rho^2(b_\rho^2-m_\pi^2) \quad \mbox{and} \quad \bar \gamma (k_p^\rho)^2 = \gamma_0^\rho + c_\rho(m_\pi/m_\pi^{\rm phys.})^2 \ ,
\end{equation}

\noindent
with $a_\rho= 0.75$ MeV, $b_\rho=480$ MeV, $c_\rho=-4.1$ MeV and $\gamma_0^\rho=44.1$ MeV.
A comparison of the fit functions and the full numerical results for the pole movements are
 shown for the $\sigma$ and $\rho$ in the left panel of 
Figs.~\ref{sigink} and \ref{rhoink}, respectively. We see
that for both $k_{\rm p}$ and $\gamma$ very simple two parameter fitting
functions provide a reasonable representation of the full results.
We start with the physical, non--vanishing values for both $\gamma$ and $k_{\rm p}$ for the
$\sigma$ as well as the $\rho$. As the pion mass gets increased
$k_{\rm p}$ decreases significantly and eventually vanishes while $\gamma$
changes relatively little.
At the point where $k_{\rm p}=0$ the two poles meet at the real axis below threshold
for the $\sigma$ and at exactly at threshold for the $\rho$, as explained
above. When the quark masses are increased further, one $\sigma$ pole moves
towards the $\pi\pi$ threshold, while the other one moves away from the
threshold
along the real $s$ (imaginary $k$) axis. 

We can now come back to the discussion of Sec.~\ref{nature}
and apply the formalism to the $\sigma$ as derived from
the IAM.
In case of the $\sigma$ the range of forces is set by $m_\rho$. The $\sigma$
becomes a bound state at  $m_\pi=450$ MeV. At this point we have
$$
\gamma \simeq \kappa \simeq 200 \ \mbox{MeV} \ \longrightarrow \ Z\simeq 0 \ . 
$$ 
Thus we conclude from this analysis that at least for $m_\pi > 450$ MeV the
$\sigma$ is predominantly of molecular nature. Note that, both for simplicity and in order to 
be conservative, we have 
shown calculations for the IAM to one-loop from \cite{Hanhart:2008mx},
although the two-loop calculation has also been performed \cite{Pelaez:2010fj}.
In that case a similar behavior is found, including the appearance of a virtual pole, although 
for pion masses $m_\pi>300$ MeV.

Given the large similarity of the pole trajectory of the $\sigma$ meson and that found for the controversial $K(800)$ scalar resonance
(or $\kappa$) with the IAM using SU(3) ChPT \cite{Nebreda:2010wv}, a similar conclusion seems unavoidable for the $K(800)$,
especially since the virtual pole predicted as the pion mass increases
was recently confirmed in a lattice-QCD calculation~\cite{Dudek:2014qha}.
%\newpage

\section{Summary}

In this paper we discussed on general grounds the properties of pole
trajectories as some strength parameter is varied for resonances coupling to
the continuum in different partial waves. There is a qualitatively different
behavior for states that couple in an $s$--wave compared to all higher
partial waves: only for $s$--wave states the two, complex conjugate resonance
poles on the second sheet meet at some value of the strength parameter below
the threshold --- for all other partial waves this meeting point is located
exactly at threshold. Using Weinberg's compositeness criterion we were able to
show that there is a connection between the value of the mentioned
subthreshold meeting point and the composition of the wave function of the
physical state.  To illustrate the mentioned properties we investigated two
models: Model A gives hadronic
molecules, which one might also call extraordinary hadrons,  for all values of the coupling that lead to a
pole, while the more general Model B
allows for a near threshold state with a prominent elementary
component, however, this requires a significant amount of fine tuning.

In lattice QCD resonance poles move as quark masses are varied. Since most
simulations at present are still performed at such values of the quark  quark masses/lattice spacings
that the resonances can not decay to the continuum, so called chiral
extrapolations are necessary to relate the lattice results to the
real world parameters. For extraordinary $s$--waves those need to contain
striking non--analyticities. This is illustrated in this paper by
employing the quark mass dependence of the $\sigma$ pole as predicted
by the inverse amplitude method in combination with one loop chiral
perturbation theory. On the basis of this study we were also able to provide
a simple parameterization for the pole trajectories that contains the
mentioned non--analyticity and should proof useful in future lattice studies.

\section*{Acknowledgements}

We are particularly thankful towards R. L. Jaffe for his participation at early stages of this work.
The research was in part supported by the Spanish project FPA2011-27853-C02-02,
DFG funds to the Sino-German CRC 110 ``Symmetries
and the Emergence of Structure in QCD'' and CRC 16 "Subnuclear Structure of Matter", as well as the EU
I3HP ``Study of Strongly Interacting Matter'' under the Seventh Framework
Program of the EU.

\end{document}